\documentclass[aip,jcp,reprint,floatfix,showpacs,amsmath,amsfonts]{revtex4-1}

\usepackage{graphicx}
\usepackage{bm}
\usepackage{array}
\usepackage{hyperref}
\usepackage{dcolumn}
\usepackage{amssymb}

\begin{document}

\title{Imaginary time Gaussian dynamics of the $\mathrm{Ar}_3$ cluster}

\author{Holger Cartarius}
\email{Holger.Cartarius@weizmann.ac.il}
\author{Eli Pollak}
\affiliation{Chemical Physics Department, Weizmann Institute of Science,
  76100 Rehovot, Israel}
\date{\today}

\begin{abstract}
  Semiclassical Gaussian approximations to the Boltzmann operator have
  become an important tool for the investigation of thermodynamic properties
  of clusters of atoms at low temperatures. Usually, numerically
  expensive thawed Gaussian variants are applied. In this article, we
  introduce a numerically much cheaper frozen Gaussian approximation to the
  imaginary time propagator with a width matrix especially suited for the
  dynamics of clusters. The quality of the results is comparable
  to that of thawed Gaussian methods based on the single-particle
  ansatz. We apply the method to the argon trimer and investigate the
  dissociation process of the cluster. The results clearly show a
  classical-like transition from a bounded moiety to three free particles
  at a temperature $T \approx 20\,\mathrm{K}$, whereas previous studies of
  the system were not able to resolve this transition. Quantum effects, i.e.,
  differences with the purely classical case manifest themselves in the
  low-temperature behavior of the mean energy and specific heat as well as
  in a slight shift of the transition temperature. We also discuss the
  influence of an artificial confinement of the atoms usually introduced to
  converge numerical computations. The results show that restrictive
  confinements often implemented in studies of clusters can influence the
  thermodynamic properties drastically. This finding may have implications
  on other studies of atomic clusters.
\end{abstract}


\maketitle

\section{Introduction}

Rare gas atomic clusters are a topic of ongoing research partially due to
the rich variety of their thermodynamic properties. Extensive studies have
been carried out on structural transformations or phase transitions 
\cite{Neirotti00a,Predescu03a,Frantsuzov04a,Predescu05a,White05a,Frantsuzov06a,Perez10a}. 
Properties of particular interest include the mean energy and specific heat.
Computations on clusters of light atoms, e.g., $\mathrm{Ne}_{13}$ and
$\mathrm{Ne}_{38}$ \cite{Predescu05a,Frantsuzov06a}, predict novel low
temperature quantum effects such as liquid-like zero temperature structures
of $\mathrm{Ne}_{38}$ as compared to a solid-like structure predicted from
classical mechanics \cite{Frantz92a}.

The features in the mean energies or specific heats of such systems appear
usually at low temperatures so that accurate quantum mechanical computational
methods are essential. Accurate calculations for multidimensional systems,
however, are challenging. Path-integral Monte Carlo methods
\cite{Berne86a,Makri99a,Ceperley03a} have been used to investigate rare gas
clusters with up to a few dozen atoms, however, they become expensive for low
temperatures so that approximations are necessary. Recently, new variants of
semiclassical initial value representations have been adopted to problems
involving the Boltzmann (imaginary time) operator $\exp(-\beta H)$. The
time evolved Gaussian method developed by Mandelshtam and coworkers
\cite{Frantsuzov03a,Frantsuzov04a} has successfully been applied to atomic
clusters
\cite{Frantsuzov03a,Frantsuzov04a,Predescu05a,Frantsuzov06a,Frantsuzov08a}
and dissipative systems \cite{Liu06a}. It is based on the imaginary time
propagation of a Gaussian wave packet of the form
\begin{multline}
  \langle \bm{x} | g \rangle = 
   \left (\pi^{3N} |\det \bm{G}(\tau)| \right )^{-1/4} \\ \times
   \exp \left ( -\frac{1}{2} [\bm{x}-\bm{q}(\tau)]^\mathrm{T}
    \bm{G}(\tau)^{-1} [\bm{x}-\bm{q}(\tau)] + \gamma(\tau) \right ) ,
  \label{eq:intro_gaussian}
\end{multline}
where for a cluster with $N$ atoms the vectors are $3N$-dimensional and
$\bm{G}(\tau)$ is a $3N\times 3N$-dimensional symmetric matrix of width
parameters.

The time evolved Gaussian method can also become expensive for high
dimensional systems. It belongs to the so-called \emph{thawed} Gaussian
methods, where the matrix of Gaussian width parameters $\bm{G}(\tau)$ changes
with time. The number of the resulting equations of motion scales with
$N^2$. This can become difficult when dealing with clusters of several dozen
atoms. Thus, Mandelshtam and coworkers introduced the single-particle ansatz
\cite{Frantsuzov04a}, in which the width matrix $\bm{G}(\tau)$ is reduced to
a block-diagonal structure by only taking into account correlations of the
coordinates of one particle and ignoring the inter-particle connections.
With this approximation one has linear scaling ($6N$) with the number of
atoms.

From a numerical point of view it is much cheaper to use \emph{frozen}
Gaussian representations of the thermal operator, in which the time-dependent
width matrix $\bm{G}(\tau)$ in Eq.\ \eqref{eq:intro_gaussian} is replaced
by a constant matrix. The number of the remaining equations of motion which
have to be solved scale as $3N$. Such a formalism has been provided by Zhang
et al. \cite{Zhang09a}. One objective of this article is to apply
the frozen Gaussian approach to the Boltzmann operator for a cluster of atoms.
We will show that by an adequate nondiagonal choice of the constant width
matrix, the frozen Gaussian method can accurately describe the mean energy,
the specific heat, and signatures of dissociation processes. To do so, we
will present a simple procedure to find a well suited shape for the width
matrix. The results we obtain are of the same quality as the single-particle
ansatz thawed Gaussian methodology, even though the frozen Gaussian variant
leaves much less freedom to the Gaussian wave packet (constant width). 

One cluster which has attracted the interest of theoretical investigations
for a long time is the argon trimer
\cite{Etters75a,Leitner89a,Leitner91a,Elyutin94a,Gonzales-Lezana99a}, whose
dissociation process has been discussed very recently in an extensive
study \cite{Perez10a}. In spite of its apparent simplicity with only three
atoms involved, the thermodynamic properties at low temperatures $T 
< 40\,\mathrm{K}$ still include open questions. In particular, 
path-integral Monte Carlo calculations of the system \cite{Perez10a} indicate
a dissociation of the three atoms at temperatures $T \gtrapprox 35\,
\mathrm{K}$ but cannot distinguish this process from structural changes. With
the semiclassical Gaussian approximations discussed in this article we are able
to provide well converged numerical mean energies and specific heats which
exhibit an unambiguous classical-like dissociation at $T \approx
20\,\mathrm{K}$. We will also address the question of how important quantum
effects are for the dissociation. Influences on the low temperature behavior
of the mean energy and the specific heat will become observable and we will
see that the transition temperature is shifted to a slightly lower value.

One focus of our discussion will be on the influence of an artificial
confinement of the atoms. For converging numerical computations of the
thermodynamic properties it is often necessary to restrict the configuration
space to a certain volume by introducing an additional confining potential
\cite{Neirotti00a}, which in practical applications usually is chosen to be
very restrictive
\cite{Neirotti00a,Predescu03a,Frantsuzov04a,Predescu05a,Frantsuzov08a}. In
this article we will show that such a restrictive choice can have a drastic
influence on the dissociation process as was already discussed in a classical
context many years ago \cite{Etters75a}. This may have implications on other
studies of atomic clusters.

The article is organized as follows. In Sec.\ \ref{sec:thermal_operator} we
introduce the Gaussian semiclassical approximation to the thermal operator.
We review the thawed Gaussian (Sec.\ \ref{sec:thawed_Gaussian}) propagator
formalism and develop a new multidimensional form of its frozen Gaussian
counterpart (Sec.\ \ref{sec:frozen_Gaussian}) capable of competing with
thawed Gaussian methods.
The results for the argon trimer based on these semiclassical Gaussian methods
are then presented in Sec.\ \ref{sec:argon_trimer}. After introducing the
system (Sec.\ \ref{sec:system}) and comparing the Gaussian methods (Sec.\
\ref{sec:results_comparison}) we discuss the influence of the confining
potential on the thermodynamic properties (Sec.\ \ref{sec:confining_sphere})
and investigate the dissociation in the classical and the quantum case
(\ref{sec:dissociation}). Conclusions are drawn in Sec.\ \ref{sec:conclusions}.

\section{Thermal operator for clusters and Gaussian approximations}
\label{sec:thermal_operator}

We consider a cluster of $N$ atoms with only internal forces depending on
the distance between the atoms, i.e., the Hamiltonian in mass scaled
coordinates has the form
\begin{equation}
  H = -\frac{\hbar^2}{2} \sum_{i=1}^N \Delta_i 
  + \sum_{j<i} V(|\bm{r}_i - \bm{r}_j|) ,
\end{equation}
where $\Delta_i$ is the Laplacian of particle $i$ and $V(|\bm{r}_i - \bm{r}_j|)$
describes the two-body interaction between particles $i$ and $j$, whose
positions are given by the vectors $\bm{r}_i$.
In the Gaussian representations used in this article it is necessary to
evaluate integrals over a product of the potential with a coherent state,
which are typically of the form 
\begin{equation}
  \langle h(\bm{q}) \rangle =  \int_{-\infty}^{\infty} d\bm{x}^{3N} \,
  \langle \bm{x} | g(\{y_i\}) \rangle^2  h(\bm{x}), 
  \label{eq:Gaverage_general}
\end{equation}
where $\langle \bm{x} | g(\{y_i\}) \rangle$ is a normalized coherent state
in $\bm{x}$, which depends on  a set of parameters $\{y_i\}$ usually
including Gaussian positions $\bm{q}$ and a width matrix $\bm{G}$, e.g.,
\begin{multline}
  \left \langle \bm{x} | g \left (\{y_i\} = \{\bm{q},\bm{G} \} \right ) 
  \right \rangle 
  = \left (\pi^{3N} |\det \bm{G}| \right )^{-1/4} \\ \times
  \exp \left ( -\frac{1}{2} [\bm{x}-\bm{q}]^\mathrm{T} \bm{G}^{-1} 
    [\bm{x}-\bm{q}] \right ) .
\end{multline}
The function $h(\bm{x})$ stands for the potential or one of its derivatives. The $3N$-dimensional
vectors $\bm{x}$ and $\bm{q}$ combine the coordinates of all $N$ atoms. It is
essential for practical applications to reduce the numerical integrations as
much as possible. Based on the facts that any central potential
can be fitted by a sum of Gaussians and that a Gaussian in the distance
$r_{ij} = |\bm{r}_i - \bm{r}_j|$ centered at the origin remains a Gaussian
in Cartesian coordinates, Frantsuzov et al. \cite{Frantsuzov04a} suggested the
implementation of the interaction potential in terms of sums of Gaussians,
\begin{equation}
  V(|\bm{r}_i - \bm{r}_j|) = \sum_{p} c_p e^{-\alpha_p r_{ij}^2} ,
  \qquad r_{ij} = |\bm{r}_i - \bm{r}_j| ,
  \label{eq:Gaussian_fit}
\end{equation}
so that all integrals of the form \eqref{eq:Gaverage_general} can be evaluated
analytically. This is of great advantage in numerical computations and is 
used for the work presented in this article.

To investigate the thermodynamic properties of the cluster we calculate the
partition function $Z(\beta)$ by evaluating Gaussian initial value
representations of the thermal operator
\begin{equation}
  K(\beta) = e^{-\beta H} ,
  \label{eq:imag_propagator}
\end{equation}
where $\beta = 1/(\mathrm{k}T)$ is the inverse temperature and 
$Z(\beta) = \mathrm{Tr}(K(\beta))$. We are interested in the mean energy
$E = \mathrm{k} T^2 \partial \ln Z/\partial T$ and the specific heat
$C = \partial E/\partial T$. In our calculations we use two different
semiclassical propagators based on a frozen and on a thawed Gaussian
representation, where in both cases the Bloch equation
\begin{equation}
  -\frac{\partial}{\partial \tau} |\bm{q}_0,\tau \rangle 
  = H  |\bm{q}_0,\tau \rangle
\end{equation}
connected with the propagator \eqref{eq:imag_propagator} is approximately
solved for a coherent state $|\bm{q}_0,\tau \rangle \approx | g(\{y_i\},\tau)
\rangle$ with either constant or variable Gaussian width parameters.

\subsection{Thawed Gaussian representation}
\label{sec:thawed_Gaussian}

The thawed Gaussian representation of the thermal operator is the most
versatile since it allows both the positions and widths of the Gaussian
wave packet to vary with time. We consider the symmetrized time evolved
Gaussian approximation (TEGA) suggested by Frantsuzov et al.
\cite{Frantsuzov03a,Frantsuzov04a},
\begin{multline}
  \langle \bm{x} | K_\mathrm{TG}(\tau) | \bm{x}' \rangle   
  = \int \frac{d\bm{q}^{3N}}{(2\pi)^{3N}} \frac{\exp[2\gamma(\tau/2)]}
  {\det[\bm{G}(\tau/2)]} \\ \times 
  \exp \left ( -\frac{1}{2} [\bm{x}-\bm{q}(\tau/2)
    ]^\mathrm{T} \bm{G}(\tau/2)^{-1} [\bm{x}-\bm{q}(\tau/2)] \right ) \\ \times
  \exp \left ( -\frac{1}{2} [\bm{x}'-\bm{q}(\tau/2)]^\mathrm{T} 
    \bm{G}(\tau/2)^{-1} [\bm{x}'-\bm{q}(\tau/2)] \right ) ,
  \label{eq:prop_TG}
\end{multline}
which is constructed from the coherent state
\begin{multline}
  \langle \bm{x} | g(\{y_i\} = \{ \bm{q}(\tau),\bm{G}(\tau) \} ) \rangle 
  = \langle \bm{x} | g(\bm{q}(\tau),\bm{G}(\tau)) \rangle \\ 
  = \left (\pi^{3N} |\det \bm{G}(\tau)|\right )^{-1/4} \\ \times
  \exp \left ( -\frac{1}{2} [\bm{x}-\bm{q}(\tau)]^\mathrm{T} \bm{G}(\tau)^{-1} 
    [\bm{x}-\bm{q}(\tau)] \right ) .
\end{multline}
One can then readily write down the partition function as:
\begin{equation}
  Z_\mathrm{TG} = \int \frac{d\bm{q}^{3N}}{(2\sqrt{\pi})^{3N}} 
  \frac{\exp[2\gamma(\tau/2)]}{\sqrt{\det[\bm{G}(\tau/2)]}} .
  \label{eq:pf_TG}
\end{equation}

The width matrix $G(\tau)$ is symmetric positive definite. The Gaussian
parameters follow the equations of motion in imaginary time $\tau$,
\begin{subequations}
  \begin{align}
    \frac{d}{d\tau} \bm{G}(\tau) &= -\bm{G}(\tau) \langle \nabla
    \nabla^\mathrm{T} V(\bm{q}(\tau)) \rangle \bm{G}(\tau) + \hbar^2 \bm{1}, 
    \label{eq:tg_eqs_motion_1} \\
    \frac{d}{d\tau} \bm{q}(\tau) &= -\bm{G}(\tau) \langle \nabla V(\bm{q}
    (\tau)) \rangle, \\
    \frac{d}{d\tau} \gamma(\tau) &= -\frac{1}{4} \mathrm{Tr} \left [ \langle
      \nabla\nabla^\mathrm{T} V(\bm{q}(\tau)) \rangle \bm{G}(\tau) \right ] 
    - \langle V(\bm{q}(\tau)) \rangle,
    \label{eq:tg_eqs_motion_3}
  \end{align}
\end{subequations}
where $\langle \dots \rangle$ represents Gaussian averaged quantities of the
form \eqref{eq:Gaverage_general}, which can be evaluated analytically for
a potential \eqref{eq:Gaussian_fit} expressed in terms of Gaussians
\cite{Frantsuzov04a}, and $\bm{1}$ is the $3N \times 3N$-dimensional identity
matrix. The boundary conditions
\begin{equation}
  \begin{aligned}
    \bm{q}(\tau \approx 0) &= \bm{q}_0 , & 
    G(\tau \approx 0) = \hbar^2 \bm{1} \tau , \\
    \gamma(\tau \approx 0) &= - V(\bm{q}_0) \tau ,
  \end{aligned}
\end{equation}
are derived by demanding that in the limit $\tau \to 0$ the Gaussian
approximation reduces to the identity operator. 

In the framework of a Gaussian propagator the thawed Gaussian representation
is usually the most accurate approximation to the exact quantum result due to
the large freedom in the parameters, as has recently been demonstrated
for a double well potential \cite{Conte10a}. However, it is also the
numerically most expensive method. The number of equations
of motion for the width matrix \eqref{eq:tg_eqs_motion_1}
scales with  $N^2$, and the matrix operations at each time step even scale
with $N^3$. This drastic increase in the required computing resources is the
most critical drawback of the method. An attempt for combining the advantages
of a thawed Gaussian propagator, where some matrix elements are still governed
by the equations of motion
\eqref{eq:tg_eqs_motion_1}-\eqref{eq:tg_eqs_motion_3}, and
avoiding the drawback of the expensive numerical effort to evaluate it, is
achieved with the so-called ``single-particle ansatz'' of Frantsuzov et al.
\cite{Frantsuzov04a}. This ansatz, or variations of it, have been applied to
several types of clusters \cite{Frantsuzov04a,Predescu05a,Frantsuzov08a}. It
uses a block-diagonal matrix $\bm{G}(\tau)$, where $3\times 3$ symmetric
matrices representing one particle along the diagonal are the only
non-vanishing matrix elements. Then the equations of motion
\eqref{eq:tg_eqs_motion_1}-\eqref{eq:tg_eqs_motion_3} are only solved for the
$3\times 3$ blocks and only the corresponding $3\times 3$ blocks of
$\langle \nabla\nabla^\mathrm{T} V(\bm{q}(\tau)) \rangle$ are included. In the
single-particle ansatz the number of equations scales with $N$ instead of
$N^2$, however, one loses information in the non-diagonal $3 \times 3$
blocks, which are set to $0$. Since the Gaussian propagators are in practical
applications usually evaluated in Cartesian coordinates, in which the motions
of the particles do not separate, important correlations between the particles
are ignored. Thus, one expects that compared to the case of a full matrix the
single-particle ansatz may lead to results of poorer quality.

In what follows we will call the full matrix variant of the thawed Gaussian
propagator FC-TG (fully coupled thawed Gaussian, also referred to as the
``fully coupled variational-Gaussian-wave-packet Monte Carlo'' in Ref.\
\onlinecite{Frantsuzov04a}) and the single-particle ansatz will be called SP-TG
(single-particle thawed Gaussian, ``single-particle
variational-Gaussian-wave-packet Monte Carlo'' in Ref.\
\onlinecite{Frantsuzov04a}).
For the argon trimer we will compare these respective approximations for the
thermodynamic properties derived from the partition function with two variants
of a frozen Gaussian propagator.

\subsection{Frozen Gaussian representation}
\label{sec:frozen_Gaussian}

The frozen Gaussian representation of the thermal operator suggested by
Zhang et al. \cite{Zhang09a} is based on a multidimensional frozen Gaussian
coherent state
\begin{multline}
  \langle \bm{x} | g(\{y_i\} = \{ \bm{p}(\tau),\bm{q}(\tau),\bm{\Gamma} \} )
  \rangle 
  = \langle \bm{x} | g(\bm{p}(\tau),\bm{q}(\tau),\bm{\Gamma}) \rangle \\ 
  = \left ( \frac{\det(\bm{\Gamma})}{\pi^{3N}} \right )^{1/4}
  \exp \biggl ( -\frac{1}{2} [\bm{x} - \bm{q}(\tau)]^\mathrm{T} \bm{\Gamma} 
    [\bm{x} - \bm{q}(\tau)] \\ +\frac{i}{\hbar} \bm{p}^\mathrm{T}(\tau)
    \cdot [\bm{x}-\bm{q}(\tau)] \biggr ) ,
\end{multline}
where $\bm{\Gamma}$ is in general a $3N \times 3N$-dimensional constant width
matrix with positive eigenvalues, and $\bm{q}(\tau)$ and $\bm{p}(\tau)$
describe the dynamical variables. The symmetrized frozen Gaussian 
approximation to the propagator reads
\begin{multline}
  \langle \bm{x}' | K_\mathrm{FG}(\tau) | \bm{x} \rangle 
  = \det(\bm{\Gamma}) \exp \left ( -\frac{\hbar^2}{4} \mathrm{Tr}(\bm{\Gamma})
    \tau \right ) \\ \times
  \sqrt{\det \left ( 2 \left [ \bm{1} - \exp (-\hbar^2 
        \bm{\Gamma} \tau) \right ]^{-1} \right )} \\
  \times \exp \left ( -\frac{1}{4} [\bm{x}' - \bm{x}]^\mathrm{T} \bm{\Gamma} 
    [\tanh(\hbar^2 \bm{\Gamma} \tau/2)]^{-1} [\bm{x}'-\bm{x}] \right ) \\
  \times \int_{-\infty}^\infty \frac{d\bm{q}^{3N}}{(2\pi)^{3N}} 
  \exp \biggl (-2 \int_0^{\tau/2} d\tau \langle V(\bm{q}(\tau)) \rangle \\
    - [\bm{\bar{x}}-\bm{q}(\tau/2)]^\mathrm{T} \bm{\Gamma} 
    [\bm{\bar{x}}-\bm{q}(\tau/2)] \biggr ) 
  \label{eq:prop_FG}
\end{multline}
with $\bm{\bar{x}} = (\bm{x}' + \bm{x})/2$ and the Gaussian averaged potential
$\langle V(\bm{q}(\tau)) \rangle$ is as defined in Eq.\
\eqref{eq:Gaverage_general}. Taking the trace yields the partition function
\cite{Zhang09a}
\begin{multline}
  Z_\mathrm{FG}(\tau) = \mathrm{Tr} \left [ K_\mathrm{FG}(\tau) \right ]
  = \sqrt{\det(\bm{\Gamma})} \exp \left ( -\frac{\hbar^2}{4}
    \mathrm{Tr}(\bm{\Gamma}) \tau \right ) \\ \times
  \sqrt{\det \left ( \left [ \bm{1}
        - \exp (-\hbar^2 \bm{\Gamma} \tau) \right ]^{-1} \right )} \\
  \times \int_{-\infty}^\infty \frac{d\bm{q}^{3N}}{(2\pi)^{N/2}} 
  \exp \left (-2 \int_0^{\tau/2} d\tau \langle V(\bm{q}(\tau)) \rangle
  \right ) .
  \label{eq:pf_FG}
\end{multline}
The numerical evaluation is relatively simple since one only needs to solve
the $3N$ imaginary time equations of motion
\begin{equation}
  \frac{\partial \bm{q}(\tau)}{\partial \tau} = -\bm{\Gamma}^{-1}
  \langle \nabla V(\bm{q}(\tau)) \rangle 
  \label{eq:FG_eqs_motion_q}
\end{equation}
for the Gaussian positions $\bm{q}(\tau)$ and only one configuration space
integration over the initial positions $\bm{q}(\tau=0)$ has to be performed.

As in the case of the SP-TG propagator the numerical scaling in the
evaluation of the two-body potential terms is $N^2$. The numerical advantage
of the frozen Gaussian propagator as compared to the FC-TG or SP-TG methods
is due to the constant width matrix, i.e., one has only to propagate the
equations of motion \eqref{eq:FG_eqs_motion_q}, whose number scales with $N$.
Additionally, functions of the width matrix $\bm{\Gamma}$ can be evaluated
in advance and do not have to be repeated at every time step since
$\bm{\Gamma}$ does not evolve in time. On the other hand, the width matrix
$\bm{\Gamma}$ is a parameter of the system and its actual choice has a
critical impact on the quality of the results \cite{Zhang09a}. It is not
trivial to find a good choice of $\bm{\Gamma}$, however, the problem
simplifies when all particles are identical.

For $N$ identical particles the simplest structure for $\bm{\Gamma}$ is a
diagonal matrix with identical width elements, i.e., a multiple of the $3N
\times 3N$ identity matrix,
\begin{equation}
  \bm{\Gamma}_1 = \Gamma \bm{1} ,
\end{equation}
with only one parameter $\Gamma$. This ansatz treats all particles equally,
is very simple to implement, and has the lowest numerical cost due to the
diagonal structure of the matrix in Cartesian coordinates. However, it ignores
the fact that a correct description of the cluster has to contain both the
free motion of the center of mass and the relative motion determined by the
particle-particle interaction potential \eqref{eq:Gaussian_fit}.

In a frozen Gaussian approximation the exact partition function of a free
particle is obtained in the limit of a vanishing Gaussian width [cf., e.g.,
Eq. \eqref{eq:pf_FG}], whereas the optimum width for the relative coordinates
can be deduced from a harmonic approximation around the minimum of the
particle-particle interaction potential and has a finite value. A separation
of the free motion of the center of mass from the internal degrees of freedom
should be avoided since it complicates the structure of the equations of
motion by introducing numerically more expensive terms such as a non-diagonal
mass matrix. Considering the thawed Gaussian propagators we note that the
FC-TG is capable of correctly describing the free center of mass motion when
it is combined with an internal potential independently of the choice of the
coordinate system, whereas this is not fulfilled for the
SP-TG\cite{Frantsuzov04a,Feldmeier00a}.

In the following we suggest a procedure based on an adequate choice of the
width matrix which allows for a correct description of the free center of
mass motion without changing the structure of the equations. To simplify, we
restrict our description to the case of three particles relevant to this
article, a generalization to an arbitrary number of particles is
straightforward. It is plausible that in a system of coordinates $\bm{R}_i$
for the center of mass
\begin{subequations}
  \begin{equation}
    \bm{R}_\mathrm{cm} = \frac{1}{3} \left ( \bm{r}_1 + \bm{r}_2 
      + \bm{r}_3 \right ) 
    \label{eq:cm_coords_1}
  \end{equation}
  and the two relative positions
  \begin{align}
    \bm{R}_1 &= \bm{r}_1 - \bm{r}_2 , \\
    \bm{R}_2 &= \bm{r}_1 - \bm{r}_3 
    \label{eq:cm_coords_3}
  \end{align}
\end{subequations}
a diagonal matrix structure is a good choice. In these coordinates the center
of mass is separated and we introduce the Gaussian width parameter $D_1$ for
its motion. The Gaussian approximation becomes exact for the center of mass
motion in the limit $D_1 \to 0$. Since all particles are equal and there is no
motivation for distinguishing between the propagation of the individual
relative coordinates, we use one single parameter $D_2$ for the remaining
coordinates. The matrix which then is applied to the coordinates
\eqref{eq:cm_coords_1}-\eqref{eq:cm_coords_3} is
\begin{equation}
  \bm{\Gamma}_\mathrm{cmc} = \begin{pmatrix}
    \bm{D}_1 & \bm{0}   & \bm{0} \\
    \bm{0}   & \bm{D}_2 & \bm{0} \\
    \bm{0}   & \bm{0}   & \bm{D}_2
  \end{pmatrix} ,
\end{equation}
where $\bm{D}_1$ and $\bm{D}_2$ are $3\times 3$ diagonal matrices with
coefficients $D_1$ and $D_2$, respectively, and $\bm{0}$ is a $3\times 3$
matrix of zeros. The most efficient way to evaluate the frozen Gaussian
partition function is to keep its structure \eqref{eq:pf_FG} in Cartesian
coordinates and to transform the width matrix into the Cartesian system
$\bm{r}_1$, $\bm{r}_2$, $\bm{r}_3$, i.e.,
\begin{equation}
  \bm{\Gamma} = \begin{pmatrix}
    (\bm{D}_1+2\bm{D}_2)/3 & (\bm{D}_1-\bm{D}_2)/3 & (\bm{D}_1-\bm{D}_2)/3 \\
    (\bm{D}_1-\bm{D}_2)/3 & (\bm{D}_1+2\bm{D}_2)/3 & (\bm{D}_1-\bm{D}_2)/3 \\
    (\bm{D}_1-\bm{D}_2)/3 & (\bm{D}_1-\bm{D}_2)/3 & (\bm{D}_1+2\bm{D}_2)/3 \\
  \end{pmatrix} .
  \label{eq:FG_full_matrix}
\end{equation}
This procedure requires the implementation of a full width matrix in the
numerical evaluation of the frozen Gaussian thermal operator but avoids a
full mass matrix and a change in the structure of the propagator. The
results for the argon trimer presented in Sec.\ \ref{sec:results_comparison}
will show that despite its simplicity this choice leads to results which are
competitive with the SP-TG propagator even though this variant of the frozen
Gaussian propagator is much cheaper to evaluate numerically. 

To distinguish the frozen Gaussian propagator with the matrix structure
of Eq.\ \eqref{eq:FG_full_matrix} from its diagonal variant we will refer in
the following to the two approximations as 2P-FG (two-parameter frozen
Gaussian) and 1PD-FG (one-parameter diagonal frozen Gaussian), respectively.

\section{The argon trimer}
\label{sec:argon_trimer}

\subsection{Atomic parameters and numerical procedure}
\label{sec:system}

To be able to compare our results with previous investigations of the argon
trimer \cite{Perez10a,Gonzales-Lezana99a} we express the pairwise interaction
by means of a Morse potential
\begin{equation}
  V(r_{ij}) = D \left ( \exp \left [ -2\alpha (r_{ij}-R_\mathrm{e}) \right ]
    - 2 \exp \left [ -\alpha (r_{ij}-R_\mathrm{e}) \right ]  \right)
  \label{eq:Morse_potential}
\end{equation}
where $r_{ij}$ is the distance between particles $i$ and $j$. The Morse
parameters are listed in Ref.\ \onlinecite{Gonzales-Lezana99a} and have been
chosen such that the Morse potential reflects a previous fit to experimental
results \cite{Aziz86a}. They are $D = 99.00\,\mathrm{cm}^{-1}$, 
$\alpha = 1.717\,\text{\r{A}}$, and $R_\mathrm{e} = 3.757\,\text{\r{A}}$.
Using the three sets of Gaussian parameters listed in Table
\ref{tab:Gaussian_parameters}
\begin{table}
  \caption{\label{tab:Gaussian_parameters}Parameters used in the Gaussian
    fit \eqref{eq:Gaussian_fit} of the Morse potential
    \eqref{eq:Morse_potential}.}
  \begin{ruledtabular}
    \begin{tabular}{lD{.}{.}{8}D{.}{.}{4}}
      $p$ & \multicolumn{1}{c}{$c_p$ [$\mathrm{cm}^{-1}$]} 
      & \multicolumn{1}{c}{$\alpha_p$ [$\text{\r{A}}^{-2}$]} \\
      \hline
      1 & 3.296\times 10^{5}  & 0.6551 \\
      2 & -1.279\times 10^{3}  & 0.1616 \\  
      3 & -9.946\times 10^{3} & 6.0600 \\
    \end{tabular}
\end{ruledtabular}
\end{table}
we achieved a very accurate Gaussian fit to this Morse potential with a
standard deviation smaller than $0.4\,\mathrm{cm}^{-1}$ in the relevant region
between $r = 3.2\,\text{\r{A}}$ and $r = 6\,\text{\r{A}}$ around the minimum.
As will become clear below this deviation is much smaller than effects due to
the Gaussian approximation used to calculate the quantum mean energy, i.e.,
differences with the previous studies of the system 
\cite{Perez10a,Gonzales-Lezana99a} do not originate from the Gaussian fit
of the potential, which is only introduced to accelerate numerical
computations.

In numerical evaluations of the partition function it is essential to
restrict the position space integrations [cf.\ $\bm{q}$ integrations in Eqs. 
\eqref{eq:pf_TG} and \eqref{eq:pf_FG}] to a reasonable region of the
configuration space containing all relevant information about the
thermodynamics of the cluster. Usually, this is done by introducing an
additional steep potential located at a certain distance $R_\mathrm{c}$
from the center of mass $\bm{R}_\mathrm{cm}$, e.g.,
\begin{equation}
  V_\mathrm{c}(\bm{r}) \propto \sum_{i=1}^{N} \left ( \frac{\bm{r}_i 
      - \bm{R}_\mathrm{cm}}{R_\mathrm{c}} \right )^{20}
\end{equation}
(cf.\ Ref.\ \onlinecite{Predescu03a}) or, as implemented in our numerics, by
restricting the sampling points $\bm{q}(\tau=0)$ in the integrations to
values $|\bm{q} - \bm{R}_\mathrm{cm}| < R_\mathrm{c}$. If the radius of the
confining sphere is chosen correctly, the restriction or its explicit form 
should have no influence on the results. The Monte Carlo sampling in
$\bm{q}$ is done with a standard Metropolis algorithm, where we followed
the procedure suggested by Frantsuzov et al. explained in detail in
Ref.\ \onlinecite{Frantsuzov04a}.

The optimum width parameter connected with the atom-atom interaction in
the two parameter ansatz \eqref{eq:FG_full_matrix} of the frozen Gaussian
propagator was found to be $D_2 = 25\,\text{\r{A}}^{-2}$ by using several
choices, monitoring the results, and comparing them to the thawed Gaussian
methods. For this choice of the parameter $D_2$ the low-temperature mean
energy reaches the smallest value, i.e., a minimum when plotted 
vs.\ $D_2$, thus, representing the best approximation to the ground
level. The observation of the smallest energy value in the limit $T
\to 0$ can be used as an additional criterion independently of the
availability of a second method such as the thawed Gaussian propagator. We note
that one can also monitor the relative amplitude of higher order corrections
to the Gaussian approximation \cite{Shao06a,Zhang09a,Conte10a} as a function of
the width parameters.

A value of $D_1 = 0.1\,\text{\r{A}}^{-2}$ is already small enough to lead to
the best possible description of the free center of mass motion. Using even
smaller values for $D_1$ did not change the results, so we decided to use
this value to have a well conditioned matrix of which the eigenvalues do not
differ too much in their order of magnitude. The parameter for a diagonal matrix
in the 1PD-FG propagator representing the best middle ground between $D_1$ and
$D_2$ is $\Gamma = 20\,\text{\r{A}}^{-2}$. It was selected with the method
described for $D_2$ above. In a previous study of the frozen Gaussian
imaginary time propagator it was found that the optimum choice for the width
parameter is almost independent of the temperature \cite{Zhang09a}, this was
also confirmed in our study of the argon trimer. We found that the radius of
the confining sphere does not have a significant influence on the optimum
choice for the width matrices. Indeed, by checking several values as
described above, it turned out that working with the same values for all
computations was the best choice.

\subsection{Comparison of the Gaussian imaginary time propagators}
\label{sec:results_comparison}

\begin{figure}
  \includegraphics[width=\columnwidth]{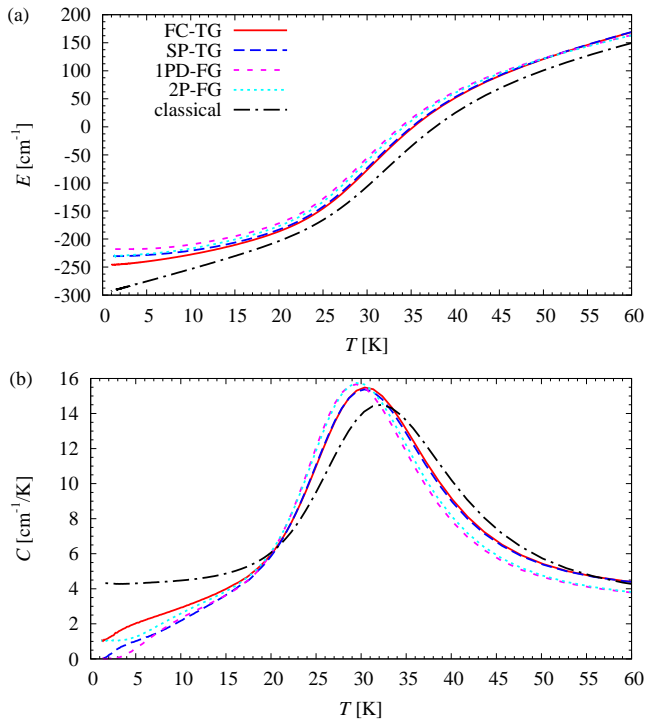}
  \caption{\label{fig:r10}Mean energy (a) and specific heat (b) of the
    argon trimer calculated for a confining radius of $R_\mathrm{c} = 10\,
    \text{\r{A}}$. Results are provided for the thawed Gaussian
    approximations FC-TG and SP-TG, the frozen Gaussian approximations
    1PD-FG, 2P-FG, and the classical theory.}
\end{figure}

\begin{figure}
  \includegraphics[width=\columnwidth]{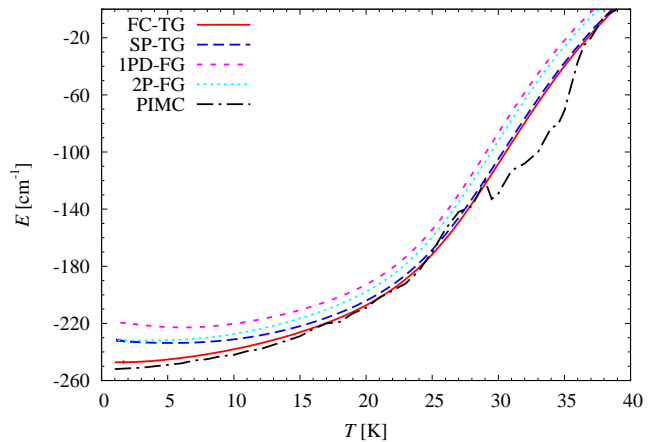}
  \caption{\label{fig:pimc_comp}Comparison of the mean energy obtained with
    the Gaussian methods FC-TG, SP-TG, 1PD-FG, and 2P-FG with the
    path-integral Monte-Carlo (PIMC) values taken from Ref.\
    \onlinecite{Perez10a}.
    The kinetic energy of the free center of mass motion is subtracted from
    our results to allow for the comparison.}
\end{figure}

First we investigated the different methods introduced in Sec.\
\ref{sec:thermal_operator} for the evaluation of the quantum partition
function for the argon trimer. To compare our results with the previous
path-integral Monte-Carlo calculation by P\'erez de Tudela et al.\
\cite{Perez10a} we selected one of their parameter sets and calculated
the mean energy and the specific heat for the argon trimer enclosed by a
confining sphere with a radius of $10\,\text{\r{A}}$, which is the weakest
confinement applied in their study. Our results obtained with the four
Gaussian propagators described above are also compared with the corresponding
derivatives of the classical partition function 
\begin{equation}
  Z_\mathrm{cl} = \left ( \frac{\mathrm{k} T}{2\pi \hbar^2} \right )^{3/2 N}
  \int e^{-\beta V(\bm{q})} \, d\bm{q}^{3N} .
  \label{eq:pf_classical}
\end{equation}
They are presented in Fig.\ \ref{fig:r10}.
To allow for the comparison with the path-integral Monte Carlo computations of
Ref.\ \onlinecite{Perez10a} in Fig.\ \ref{fig:pimc_comp}
we subtracted for this figure the exact kinetic energy of the free
center of mass $E_\mathrm{cm} = 3/2 \mathrm{k}T$ from our values, which, in
the calculation, always include the energy of the whole cluster including
the center of mass translation.

The four semiclassical methods are in reasonably good agreement with each
other and the classical results, however, there are quantitative differences.
One can expect that the thawed Gaussian imaginary
time propagator with a full matrix (FC-TG) provides the best approximation to
the exact quantum results, and indeed the mean energy obtained with that method
shows the best correspondence with the path-integral Monte Carlo computations
of Ref.\ \onlinecite{Perez10a} (cf.\ Fig.\ \ref{fig:pimc_comp}). In
particular, in the low-temperature limit the FC-TG propagator gives the best
approximation ($E = -246\,\mathrm{cm}^{-1}$) to the ground state
energy\cite{Perez10a} of $-252.44\, \mathrm{cm}^{-1}$.
Even though the FC-TG method is the most versatile propagator applied
here, it still assumes a Gaussian form of the atom's wave functions
and shows a slight difference as compared to the numerically exact
path-integral Monte Carlo ground state energy.
For higher temperatures there appear larger deviations as compared to the
path-integral Monte Carlo results of Ref.\ \onlinecite{Perez10a}, however, the
latter ones are not very accurate due to the larger statistical error with
increasing temperature.
The semiclassical Gaussian approximations are expected to describe the 
partition function and its derivatives at higher temperatures even better
than at low temperatures since they converge to the correct answer in the
classical limit. We can thus assume that our results are more accurate as the
temperature increases. The differences between our results and those of
Ref.\ \onlinecite{Perez10a} at higher temperatures $T \gtrapprox 30\,\mathrm{K}$
become even clearer when considering the specific heat. Our Gaussian
calculations indicate a direct transition from the bound cluster to the
completely dissociated situation with three free atoms, whereas such a clear
conclusion is not possible with the path-integral Monte Carlo calculations
of Ref.\ \onlinecite{Perez10a}. This transition is discussed in more detail in
the next two sections.

It is also expected that the worst approximation in the low-temperature limit
is obtained by the 1PD-FG ansatz. As was mentioned in Sec.\
\ref{sec:frozen_Gaussian} the diagonal width matrix does not treat the free
center of mass motion correctly. This manifests itself as a large deviation
of the mean energy from the correct value for $T \to 0$. The energy calculated
with the diagonal frozen Gaussian width matrix increases for temperatures 
below $5\,\mathrm{K}$, and this is definitely wrong. This deficiency is
overcome with the 2P-FG matrix as presented in Eq.\ \eqref{eq:FG_full_matrix}.
It leads to a considerably lower mean energy in the low-temperature limit,
which is closer to the exact ground level and is also closer to the FC-TG
energy values, which can be considered to provide the best values of all
methods used here. 

More interesting, however, is the comparison of the 2P-FG and the SP-TG
propagators. As reported in previous investigation of larger clusters
\cite{Frantsuzov04a} the single-particle thawed Gaussian approximation shows
results for the mean energy and the specific heat which are qualitatively in
agreement with the full matrix case. Quantitative differences have
been reported and can also be found in our calculations for temperatures $ T
\lessapprox 45\, \mathrm{K}$. The differences with respect to the full matrix
results increase with decreasing temperature. However, as can be seen in Fig.\
\ref{fig:r10}(a) the description of the mean energy of the SP-TG and the 2P-FG
methods is of similar quality at very low temperatures.
This is a remarkable finding since the evaluation of the single-particle
thawed Gaussian propagator is more expensive than the frozen Gaussian due
to the need to propagate the width matrix elements in time.

\begin{figure}
  \includegraphics[width=\columnwidth]{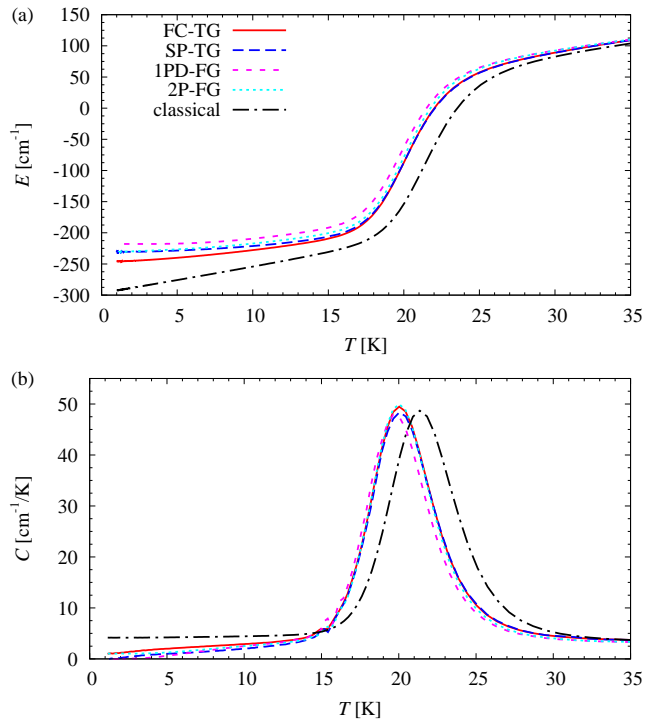}
  \caption{\label{fig:r32}Mean energy (a) and specific heat (b) of the
    argon trimer calculated with the Gaussian approximations  FC-TG, SP-TG,
    1PD-FG, 2P-FG, and classically for $R_\mathrm{c} = 32\, 
    \text{\r{A}}$.}
\end{figure}

By contrast, the specific heat curve of the 2P-FG propagator is
closer to that of the 1PD-FG than to the two thawed Gaussian approximations
which agree very well with each other. It seems that in the temperature
region around the transition from the bound to the dissociated cluster, a
thawed Gaussian can provide a better description. However, the deviation
is small and vanishes for weaker external confinements. An example for
such a weaker confinement is shown in Fig.\ \ref{fig:r32},
in which the comparison of Fig.\ \ref{fig:r10} is repeated for a confinement
of $R_\mathrm{c} = 32\, \text{\r{A}}$. Here, down to a temperature of
$18\,\mathrm{K}$ the two thawed Gaussian approximations and the 2P-FG
propagator provide almost identical results. The lines lie on top of each other
both for the mean energy and the specific heat. Only the 1PD-FG values
deviate a bit from the three other methods. 

In summary, we can state that the thawed Gaussian propagator with a full matrix
provides the best approximation which deviates in the low-temperature limit
only slightly from the exact result and almost reaches the ground level for
$T \to 0\,\mathrm{K}$. In all cases in which the high accuracy of the full
matrix thawed Gaussian propagator is not required or in which the integration
of the equations of motion for the width parameters of a full matrix is too
expensive, the frozen Gaussian ansatz with two parameters (2P-FG) seems to be
the best choice. It provides the same quality of results as the
single-particle thawed Gaussian propagator but is much easier to evaluate
since no integrations of parameters of the width matrix are required.

\subsection{Dissociation and the influence of the confining sphere}
\label{sec:confining_sphere}

\begin{figure}
 \includegraphics[width=\columnwidth]{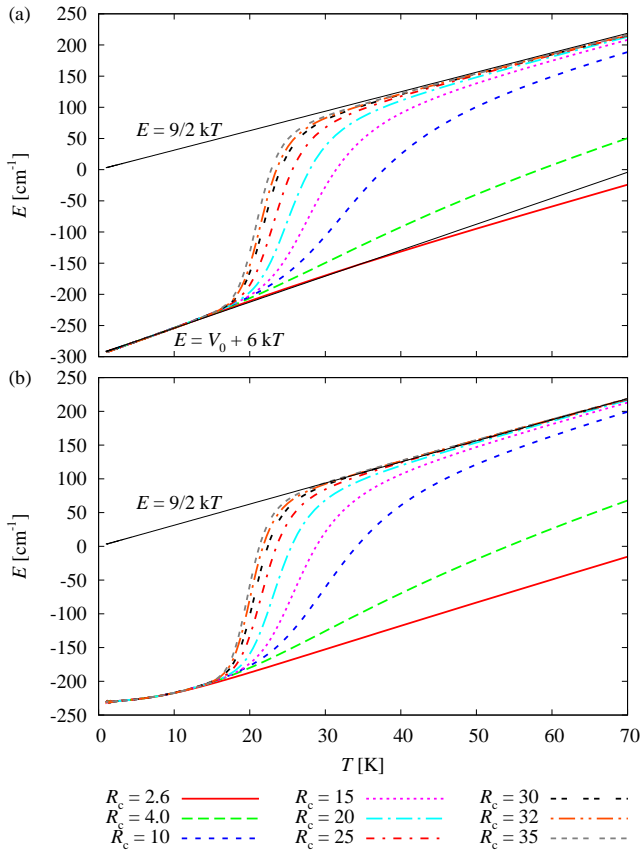}
  \caption{\label{fig:r_values}Comparison of the mean energy of the argon
    trimer for different confining radii $R_\mathrm{c}$ (in \r{A}) in
    classical (a) and quantum calculations with the 2P-FG partition
    function (b). We also added the thin black lines $E = 9/2\mathrm{k}T$
    representing three free particles and $E = V_0 + 6\mathrm{k}T$ with
    the potential minimum $V_0 = -296\, \mathrm{cm}^{-1}$ for the classical
    low-temperature behavior.}
\end{figure}

Before discussing the dissociation process in more detail we have to
investigate the influence of the confining sphere on the mean energy and
the specific heat. While often a very restrictive value of the confining
radius $R_\mathrm{c}$ is
chosen\cite{Neirotti00a,Predescu03a,Frantsuzov04a,Predescu05a,Frantsuzov08a},
our calculations demonstrate that its value significantly affects the
thermodynamic properties of the clusters unless large values, drastically
exceeding a few \r{A}ngstr\"{o}ms, are used. Figure \ref{fig:r_values}
shows the mean energy of the argon trimer for several confinements
$R_\mathrm{c}$ in a purely classical calculation [Fig.\ \ref{fig:r_values}(a)]
and an evaluation of the quantum partition function with the 2P-FG propagator
[Fig.\ \ref{fig:r_values}(b)], which, as was discussed in Sec.\ 
\ref{sec:results_comparison}, describes the quantum behavior correctly.
The values $R_\mathrm{c} = 2.6\,\text{\r{A}}, 4\, \text{\r{A}}$, and
$10\,\text{\r{A}}$ have already been used in Ref.\ \onlinecite{Perez10a}. The
qualitative behavior of these three curves differs strongly. In particular,
the mean energy for $R_\mathrm{c} = 10\,\text{\r{A}}$ shows a significant
increase of the slope for temperatures above $20\,\mathrm{K}$. In Ref.\
\onlinecite{Perez10a} this was regarded as an indication that this radius
allows for a total fragmentation of the cluster. However, the mean energy for
higher temperatures reveals that this is not fulfilled completely. We included
the line for $E = (9/2)\mathrm{k} T$ in the figure, which corresponds to three
free particles. The mean energy for the restriction $R_\mathrm{c} = 10\,
\text{\r{A}}$ does not reach that line even at $T = 70\,\mathrm{K}$. However,
for the weaker restrictions $R_\mathrm{c} = 15\dots 35\,\text{\r{A}}$ one
can observe that actually a total fragmentation takes place and that for
$T \gtrapprox 40\,\mathrm{K}$ the energy is identical to that of three free
particles.
In Fig.\ \ref{fig:r_values}(a) we also included the line $E = V_0 +
6\mathrm{k}T$  with the potential minimum $V_0 = -296\,\mathrm{cm}^{-1}$.
This line corresponds to the classical expectation at low temperatures for
the internal rotations and oscillations of the trimer plus the energy of the
free center of mass. 

Even though the behavior at very low ($T \leq 15\, \mathrm{K}$) and high
temperatures ($T > 35\,\mathrm{K}$) agrees well for all confinements
$R_\mathrm{c} > 15\,\text{\r{A}}$ the transition itself obviously depends more
critically on the value of $R_\mathrm{c}$. It is clear that a confining radius
$R_\mathrm{c} = 10\,\text{\r{A}}$ is too restrictive and does not describe the
cluster correctly. A significantly larger confining radius $R_\mathrm{c} >
30\,\text{\r{A}}$ is required. For the largest confining radii used in
Fig.\ \ref{fig:r_values} we observe convergence, i.e., a further expansion of
the confining sphere does not change the results significantly. 

We note that it becomes increasingly difficult to converge the Monte-Carlo
integrations for the classical and the 2P-FG partition function for increasing
$R_\mathrm{c}$. Similarly, the large error bars in the path-integral Monte Carlo
calculation of Ref.\ \onlinecite{Perez10a} indicate that in their computations a
radius of $R_\mathrm{c} = 10\,\text{\r{A}}$ was already challenging.
Nevertheless, the results presented in Fig.\ \ref{fig:r_values} demonstrate
that the added effort of increasing $R_\mathrm{c}$ beyond $30\,\text{\r{A}}$
is essential. The necessity for a thorough investigation of the correct
boundary conditions is already known from classical investigations of
atomic clusters. Etters and Kaelberer \cite{Etters75a} demonstrated the
negative influence of too restrictive boxes on the classical average energy.

\subsection{Dissociation from classical and quantum perspectives}
\label{sec:dissociation}

The cluster at a confinement of $R_\mathrm{c} = 32\,\text{\r{A}}$ can
be regarded as converged with respect to $R_\mathrm{c}$. The artificial
confinement does not have a further noticeable influence on the thermodynamic
properties. This allows us to discuss the features observed in the mean
energy and the specific heat in more detail. Additionally, the various
Gaussian propagators used to obtain the quantum properties agree with each
other to a high precision, so that we may consider them as converged in the
sense that the choice of Gaussian method has no further influence.

As can be seen in Fig.\ \ref{fig:r32} the qualitative behavior in the quantum
and classical cases is almost the same. The cluster is bound at low
temperatures, shows a relatively sharp transition, and is completely
dissociated for temperatures $T > 33\,\mathrm{K}$. For very low temperatures
the classical mean energy exhibits the expected behavior $E \propto
6\mathrm{k}T$ [cf.\ also Fig.\ \ref{fig:r_values}(a)] for the system (free
center of mass, rotations and oscillations of the internal degrees of
freedom), whereas the FC-TG mean energy (best
approximation, see Sec.\ \ref{sec:results_comparison}) converges to
a value close to that of the ground level. Further differences between the
classical and quantum mechanical results are found in the transition
region. It is shifted to slightly lower temperatures in the quantum
calculations as compared to the classical. All quantum calculations
show a maximum of the specific heat at $20\,\mathrm{K}$, while the classical
maximum is at $21.5\,\mathrm{K}$. One reason for this shift is the
presence of the zero point energy in the quantum system. As is already obvious
in Fig.\ \ref{fig:r_values} the classical and quantum results agree very well
in the high-temperature limit, since both trend to the case of three free
particles. Thus, we conclude that the transition observed in the cluster of
three argon atoms is a classical phenomenon. The only difference between
classical and quantum mechanics is in the temperature at which the transition
occurs. We note that Etters and Kaelberer \cite{Etters75a} reported a
``liquid-gas transition'' identified by the absence of bounded atom
configurations in a classical investigation of the system with
``free-surface boundary conditions'', i.e., without a confining sphere, at
$T = 20\,\mathrm{K}$, which is in good agreement with our results.

Our finding of a classical-like complete dissociation of the trimer in one
step differs from the conclusions of  P\'erez de Tudela et al.\
\cite{Perez10a}. While the mean energy in the calculations of
Ref.\ \onlinecite{Perez10a} for $R_\mathrm{c} = 10\,\text{\r{A}}$ shows a
larger and larger slope for increasing temperatures up to $T = 40\,\mathrm{K}$
we observe already a decrease of the slope for temperatures $T >
30\,\mathrm{K}$. The difference relative to the path-integral Monte Carlo
calculations becomes even more pronounced in the specific heat. P\'erez de
Tudela et al.\ \cite{Perez10a} report that they find an ``apparent" maximum
which evolves with the radius $R_\mathrm{c}$ of the confinement and appears
slightly below $40\,\mathrm{K}$ for $R_\mathrm{c} = 10\,\text{\r{A}}$.
The absence of an unambiguous maximum was seen as an indication for
structural changes of the cluster instead of a proper ``phase transition''.
Although the dissociation of the cluster is not fully achieved for such a
strong confinement, it is clear from the Gaussian methods used in this
article that already for $R_\mathrm{c} = 10\,\text{\r{A}}$ a pronounced peak
in the specific heat indicating the dissociation of the system at $T
\approx 30\, \mathrm{K}$ is present (cf.\ Fig.\ \ref{fig:r10}). Describing
the cluster with a weaker confinement correctly reveals the unambiguous
dissociation without intermediate structural modification as discussed above.

The very low temperature found for the dissociation of the cluster may also be
important for rare gas clusters in general. As was already
discussed\cite{Perez10a}, even for a confining sphere with $R_\mathrm{c}
= 10\,\text{\r{A}}$, the transition temperature of $T \approx 35\,\mathrm{K}$
is lower than temperatures discussed for structural transformations or a
``melting'' of clusters. Features indicating such changes have, e.g., been
found beyond $40\, \mathrm{K}$ for argon \cite{Pahl08a}. If one considers that
the dissociation temperature is actually even lower ($T\approx
20\,\mathrm{K}$, c.f.\ Fig.\ \ref{fig:r32}), one necessarily concludes that it
is wrong to ignore the influence of the confining sphere on such properties.
In larger neon clusters ($\mathrm{Ne}_{13}$ and $\mathrm{Ne}_{38}$) features
in the mean energy or the specific heat which were related to structural
changes, have been reported between $6\,\mathrm{K}$ and $8\,\mathrm{K}$
\cite{Neirotti00a,Frantsuzov04a,Predescu05a}. These temperatures are lower
than the dissociation found here, however, one may expect that at least a
partial dissociation can set in much earlier in larger clusters since they
contain higher energies. In the numerical simulations
\cite{Neirotti00a,Frantsuzov04a,Predescu05a} the confining radii are chosen
such that no atom can leave the cluster during the time evolution. Our
results indicate that such a constraint might be too restrictive and lead to
incorrect conclusions. A partial or full dissociation can influence structural
transformations and may even set in before structural changes of an
artificially confined cluster can occur.

\section{Conclusions and outlook}
\label{sec:conclusions}

In the present article we investigated the argon trimer by means of
semiclassical Gaussian approximations to the Boltzmann operator. We introduced
a new matrix structure for a frozen Gaussian variant of the imaginary time
propagator which is capable of correctly dealing with the free center of
mass motion of a cluster of atoms in Cartesian coordinates. With this matrix
structure we were able to show that the frozen Gaussian propagator is,
in spite of its simplicity, competitive with numerically more expensive
thawed Gaussian variants. In particular, the frozen Gaussian method provides
the same quality thermodynamic results as the so-called single-particle
thawed Gaussian propagator, which in addition to the time-dependent variables
of the frozen Gaussian requires the time evolution for the elements of a
block-diagonal width matrix. This is especially true in the low-temperature
limit, where quantum effects become important and the form of the
semiclassical approximation is supposed to have the largest influence. Our
results suggest that the frozen Gaussian ansatz with two parameters (2P-FG)
introduced in this article is the method of choice in all cases in which the
higher accuracy of the full matrix thawed Gaussian propagator is not required
or in which the integration of the equations of motion for the width
parameters of a full matrix is too expensive.

The evaluation of the mean energy and the specific heat for the cluster of
three argon atoms demonstrated that a previous investigation of the system
\cite{Perez10a} used too restrictive confinements to describe
the dissociation behavior of the system correctly. Above $T = 15\,\mathrm{K}$
the cluster directly dissociates into three free atoms, as is evident from 
the Gaussian calculations presented in this article. This dissociation is
almost purely classical, the influence of quantum mechanics is only found
in a convergence to the ground state energy instead of the classical potential
minimum for $T \to 0$ and in a slight shift of the three-body dissociation
transition to higher classical temperatures ($\Delta T \approx
1.5\,\mathrm{K}$) which we attribute to the zero point energy of 
quantum mechanics. 
The clear and pronounced transition found in this article supports
the conclusion of Ref.\ \onlinecite{Perez10a} that the dissociation of
the atoms from the cluster is important when reconfigurations of the
internal structure are considered.
Our results strongly indicate that the confinement to very small spheres
usually applied in the calculation of the partition function and values
deduced from it \cite{Predescu03a,Frantsuzov04a,Predescu05a,Frantsuzov08a}
might be too restrictive to fully understand the low-temperature behavior
of the clusters. The dissociation can set in before structural changes
or a melting can be observed. To make a clear statement on this question
it is necessary to advance the investigations done here to clusters with
higher numbers of atoms. In particular, the cases of 
$\mathrm{Ar}_{6}$ \cite{Franke88a}, $\mathrm{Ar}_{13}$ 
\cite{Franke88a,Borrmann94a,Tsai93a}, $\mathrm{Ne}_{13}$ \cite{Frantsuzov04a}
or $\mathrm{Ne}_{38}$ \cite{Predescu05a} examined recently are of special
interest.

On the technical side, it is known that both the frozen and thawed Gaussian
propagators used here are, in the framework of a generalized time-dependent
perturbation theory \cite{Zhang03a,Pollak03a}, the lowest order
approximations in a series converging to the exact quantum propagator
\cite{Shao06a,Zhang09a,Conte10a}. Higher orders can help to understand the
thermodynamic properties better and to verify the results obtained here
with a higher accuracy. Furthermore, the corrections obtained by
the evaluation of higher order terms provide objective access to the quality
with which the Gaussian approximations reflect the quantum effects in the
system studied. They are the topic of current studies.

\begin{acknowledgments}
H.C. is grateful for a fellowship from the Minerva Foundation. This work
was supported by a grant of the Israel Science Foundation.
\end{acknowledgments}

\end{document}